\documentclass[]{article}
\usepackage{amssymb,latexsym,amsmath,enumerate,verbatim,amsfonts,amsthm}
\usepackage{cite}
\title{Boundedness from below conditions for a general scalar potential of two real scalars fields and the Higgs boson \footnote{This  work was supported by
		the National Natural Science Foundation of P.R. China (Grant No.12171064), by The team project of innovation leading talent in chongqing (No.CQYC20210309536) and by the Foundation of Chongqing Normal university (20XLB009)}}
\author{Yisheng Song$^1$, Liqun Qi$^{2,3}$\\
$^{1}$  \small  School of Mathematical Sciences, Chongqing Normal University, \\
\small Chongqing, P.R. China, 401331; Email: yisheng.song@cqnu.edu.cn\\
$^{2}$ \small Department of Mathematics, School of Science, \\
\small Hangzhou Dianzi University, 
Hangzhou 310018, P. R. China; \\
$^{3}$ \small Department of Applied Mathematics, The Hong Kong\\ \small Polytechnic University, 
Hung Hom, Kowloon, Hong Kong;\\ \small Email: maqilq@polyu.edu.hk}

\begin{document}

\maketitle

\begin{abstract}
The most general scalar potential of two real scalar fields and a Higgs boson is a quartic homogeneous polynomial  about 3 variables, which defines a 4th order 3 dimensional symmetric tensor. Hence, the boundedness from below of such a scalar potential involves the positive (semi-)definiteness of the corresponding tensor. So, we mainly discuss analytical expressions of positive (semi-)definiteness for such a special 4th order 3-dimension symmetric tensor in this paper. Firstly, an analytically necessary and sufficient condition  is given to test the positive (semi-)definiteness of a 4th order 2 dimensional symmetric tensor. Furthermore, by means of such a result, the necessary and sufficient conditions of  the  boundedness from below are obtained for a general scalar potential of two real  scalar fields  and the Higgs boson. 

\end{abstract}

\section{Introduction}
The boundedness from below  of a scalar potential makes physical sense, which simply implies that such a scalar potential is positive ( or non-negative). The polynomial degree of the potential is $4$ when one keeps the scalar interactions renormalizable \cite{IKM2018}. Then the condition for the potential of $n$ real scalar fields $\phi_i$ ($i=1,2,\cdots,n$) to be bounded from below in the strong sense is equivalent to the requirement that 
\begin{equation}\label{eq:1} V(\phi) =\sum_{i,j,k,l=1}^n v_{ijkl}\phi_i\phi_j\phi_k\phi_l>0. \end{equation}
Let $\mathcal{V}=(v_{ijkl})$. Then $\mathcal{V}$ is a 4th order symmetrical tensor, and hence, the above requirement \eqref{eq:11} is the positive definiteness of the tensor $\mathcal{V}$.  Qi \cite{LQ1,LQ5} first used and introduced  the positive definiteness and copositivity of tensors. An  $m$th order $n$ dimensional real tensor $\mathcal{V}=(v_{i_1i_2\cdots i_m})$ is said to be \begin{itemize}
	\item[(i)] {\em positive semi-definite} if  $\mathcal{V}x^m=\sum\limits_{i_1,i_2,\cdots,i_m=1}^nv_{i_1i_2\cdots
		i_m}x_{i_1}x_{i_2}\cdots
	x_{i_m}\geq0$ for all $x\in \mathbb{R}^n$ and an even number $m$;
	\item[(ii)] {\em positive definite} if  $\mathcal{V}x^m>0$ for all $x\in \mathbb{R}^n\setminus\{0\}$ and an even number $m$;
	\item[(iii)] {\em copositive} if  $\mathcal{V}x^m\geq0$ for all $x\geq 0$;
	\item[(iv)] {\em strictly copositive} if  $\mathcal{V}x^m>0$ for all $x\geq 0$ and $x\ne 0$.
\end{itemize}

Kannike \cite{K2016,K2018,K2012} presented the vacuum stability conditions of general scalar potentials
of  two real scalar fields $\phi_1$ and $\phi_2$ and the Higgs bonson $\mathbf{H}$, and studied the sufficient condition of boundedness from below for scalar  potential of the {\bf Standard Model} ({for short, \bf SM}) Higgs $\mathbf{H}_1$, an inert doublet $\mathbf{H}_2$ and a complex singlet $\mathbf{S}$. In fact, such two problems were solved by Kannike \cite{K2016}, where the  first problem involves  the positive definiteness of the corresponding symmetric tensor and the  second problem requires the copositivity of the corresponding  symmetric tensor.   Chauhan \cite{GC} gave an analytical  vacuum stability condition of the left-right symmetric model  for successful symmetry breaking. Ivanov \cite{IF2020}  presented  the stability conditions in multi-Higgs potentials.  Bahl et.al. \cite{BCCIW} provided  the analytically sufficient conditions of the vacuum stability for the  two-Higgs-doublet potential with CP conservation, and showed a vacuum stability condition for the  two-Higgs-doublet  potential with CP violation depends on the Lagrange multiplier $\zeta$.  Recently, Song \cite{S2022} established  the boundedness from below conditions of scalar potential for the two-Higgs-doublet  with explicit CP conservation.  Song \cite{S2023} obtained  the boundedness from below conditions of scalar potential for a general  two-Higgs-doublet, which includes  necessary conditions and sufficient conditions.  Also see Faro-Ivanov \cite{FI2019}, Belanger-Kannike-Pukhov-Raidal \cite{BKPR,BKPR2014}, Ivanov- K\"{o}pke-M\"{u}hlleitner \cite{IKM2018} for more details. In Refs. \cite{IF2020,MR2001,IV2012,IV2013,IKO2010},  one can construct only one quadratic term and five quartic terms for the Higgs potential with the help of three Higgs doublets with equal electroweak quantum numbers, which is a quartic polynomial with real coefficients defined on complex field. Toorop-Bazzocchi-Merlo-Paris \cite{ABMP2011,ABMP2013} and Degee-Ivanov-Keus \cite{DIK2013} turned such a  polynomial from complex field to real field. In fact, they were trying to look for the analytical condition of such a polynomial to be positive.

Recently, Song-Qi \cite{SQ2019} and Liu-Song \cite{LS2019}  gave a different  sufficient condition of copositivity for 4th order 3 dimensional symmetric tensors to find the boundedness from below conditions of scalar  potential of the {\bf SM} Higgs $\mathbf{H}_1$, an inert doublet $\mathbf{H}_2$ and a complex singlet $\mathbf{S}$. Very recently, Qi-Song-Zhang \cite{QSZ2020} presented a necessary and sufficient condition of copositivity for such a tensor given by the above particle physical model. Song-Li \cite{SL2022} provided an analytically necessary and sufficient condition of the boundedness from below for such a scalar  potential model.

In the past decades, many numerical algorithms were established to find  some H-(Z-)eigenvalues of a tensor \cite{LQ1,NQW2008,NQZ2009,NZ2015,HLQS2013,ZQLX2013,HCD2015,HCD-2015,H2013,CDN2014,CHZ2016,CW2018,QCC2018,QL2017},  and  were applied to test the positive definiteness of such an even order tensor by means of the sign of the smallest H-(Z-)eigenvalue. On the other hand, some classes of  tensors with special structure may be determined directly their positive definiteness such as Hilbert tensor \cite{SQ2014},  diagonal dominant tensor \cite{LQ1}, B-tensor \cite{S-Q2015,QS2014,LQL2015} and others.
However, the practical matters such as the vacuum stability of general scalar potentials of a few fields require analytical expressions. The most general scalar potential of two real scalar fields $\phi_1$ and $\phi_2$ and the Higgs doublet $\mathbf{H}$ (Kannike \cite{K2016,K2018,K2012})  is
\begin{equation}\label{eq:2}
	\begin{aligned}
		V(\phi_1,\phi_2,|H|)=& \lambda_{H}|H|^4+\lambda_{H20}|H|^2\phi_1^2+\lambda_{H11}|H|^2\phi_1\phi_2+\lambda_{H02}|H|^2\phi_2^2\\
		&\ +\lambda_{40}\phi_1^4+\lambda_{31}\phi_1^3\phi_2+\lambda_{22}\phi_1^2\phi_2^2+\lambda_{13}\phi_1\phi_2^3+\lambda_{04}\phi_2^4.
	\end{aligned}
\end{equation}
Clearly, such a quartic homogeneous polynomial  defines a 4th order 3 dimensional symmetric tensor $\mathcal{V} = (v_{ijkl})$, \begin{equation}\label{eq:3}
	\begin{aligned}v_{1111}=&\lambda_{40},\  v_{2222}=\lambda_{04},\  v_{3333}=\lambda_{H},\ v_{1112}=\frac14\lambda_{31},\ v_{1222}=\frac14\lambda_{13},\\
		v_{1133}=&\frac16\lambda_{H20},\ v_{1122}= \frac16\lambda_{22},\ v_{2233}=\frac16\lambda_{H02},\\
		v_{1233}=&\frac1{12}\lambda_{H11},\ \  v_{ijkl}=0\mbox{ for the others}.
\end{aligned}\end{equation}
and hence, the boundedness from below of such a scalar potential involves the positive (semi-)definiteness of such a tensor $\mathcal{V} $. So this requires an analytic condition of positive (semi-)definiteness. For a  4th order 2 dimensional symmetric tensor, the analytical condition of the positive definiteness traced back to ones of  Refs. Gadem-Li \cite{GL1964}, Ku \cite{K1965} and Jury-Mansour \cite{JM1981}. Wang-Qi \cite{WQ2005} improved their proof and conclusions.  However, the above result depends on the discriminant of such a polynomial. Recently, Guo \cite{G2020} showed a new necessary and sufficient condition without the discriminant. Very recently, Qi-Song-Zhang \cite{QSZ20202} gave a new necessary and sufficient condition other than the above results.  Hasan-Hasan \cite{HH1996} claimed that a necessary and sufficient condition of positive definiteness  was proved without the discriminant. However, there is a problem in their argumentations. In 1998, Fu \cite{F1998} pointed out that Hasan-Hasan's results are sufficient only. Song \cite{S2021} gave several analytically  sufficient conditions of the positive definiteness of 4th order 3 dimensional symmetric tensor. Until now,  peoples have  not  found an analytically necessary and sufficient condition  of positive definiteness for a 4th order 3 dimensional symmetric tensor.

In this paper, we mainly concentrate on the analytical expressions of positive definiteness for a special 4th order tensor given by \eqref{eq:13}. More precisely, by means of Qi-Song-Zhang's result,  we first show  an analytically necessary and sufficient condition of positive (semi-)definiteness of 4th order 2 dimensional symmetric tensors. Secondly, with the help of this conclusion, we discuss positive (semi-)definiteness of a 4th order 3-dimension symmetric tensor defined by \eqref{eq:13}. Then these analytic conditions are the necessary and sufficient conditions of the boundedness from below for a scalar potential \eqref{eq:12} of two real scalar fields $\phi_1$ and $\phi_1$ and the Higgs doublet $\mathbf{H}$.
\vskip 4mm
\section{\bf 4th order  symmetric real tensor}
\vskip 4mm
A 4th order 3 dimensional real tensor $\mathcal{V}$ consists of $81$ entries in the real field $\mathbb{R}$, i.e.,
$$\mathcal{V} = (v_{ijkl}),\ \ \ \ \  v_{ijkl} \in \mathbb{R},\ \  i,j,k,l=1,2,3.$$
A  tensor $\mathcal{V}$ is said to be {\em symmetric} if its entries $v_{ijkl}$ are invariant for any permutation of its indices. It is well-known that a 4th order 3 dimensional symmetric tensor $\mathcal{V}$ is composed of $15$ independent entries only,
\begin{equation}\label{eq:4}\begin{aligned}
		&v_{1111}, v_{2222},\ v_{3333},\ v_{1222},\ v_{1333},\ v_{1112},\ v_{1113},\ v_{2333},\\
		&v_{2223},\ v_{1122},\ v_{1133},\ v_{2233},\ v_{1223},\ v_{1123},\ v_{1233}.
\end{aligned}\end{equation}
A 4th order 2 dimensional symmetric tensor $\mathcal{V}$ is composed of $5$ independent entries only,
$$ v_{1111}, v_{2222},\  v_{1222},\ v_{1112},\ v_{1122}.$$
It is obvious that there is  a consistent one-to-one match between a 4th order 3 dimensional symmetric tensor and a quartic homogeneous polynomial  with 3 variables.
Such a homogeneous polynomial, denoted as $\mathcal{V}x^4$, i.e.,
\begin{equation}\label{eq:5}\mathcal{V}x^4=\sum_{i,j,k,l=1}^3v_{ijkl}x_ix_jx_k
	x_l.\end{equation}

Let $\|\cdot\|$ denote any norm on $\mathbb{R}^n$. Then the following conclusions on unit sphere are known \cite{LQ1,QL2017,QCC2018}.\\
\\
Let $\mathcal{V}$ be a 4th order symmetric tensor and let $S$ be the unit sphere on $\mathbb{R}^n$, $S=\{x\in \mathbb{R}^n: \ \|x\|=1 \}$. Then
\begin{itemize}
	\item[(i)] $\mathcal{V}$ is positive semi-definite if and only if $\mathcal{V}x^4\geq0$ for all $x\in S$;
	\item[(ii)] $\mathcal{V}$ is positive definite if and only if $\mathcal{V}x^4>0$ for all $x\in S$.
\end{itemize}
\vskip 4mm
\section{\bf Non-negativeness of quadratic and quartic polynomial}
\vskip 4mm
Let  $P_2(t)$ be a quadratic polynomial,
\begin{equation}\label{eq:6}
	P_2(t) = at^2 + bt + c,
\end{equation}
with $a>0$.  Then  the following results should be well-known, which is showed hundreds of years ago. Also see Qi-Song-Zhang \cite{QSZ20202}.
\vskip 2mm
(1) $P_2(t) > 0$  for all $t \ge 0$ if and only if
\begin{equation}\label{eq:7}	
	b \ge 0\mbox{ and }c > 0;\ \
	b < 0\mbox{ and }4ac - b^2>0.
\end{equation}

(2) $P_2(t)  \ge 0$ for all $t \ge 0$ if and only if
\begin{equation}\label{eq:8}	
	b \ge 0\mbox{ and }c \ge 0;\ \
	b < 0\mbox{ and }4ac - b^2 \ge 0.
\end{equation}
\vskip 4mm

Let $P_4(t)$ be a quartic polynomial,
\begin{equation}\label{eq:9}	P_4(t) = at^4 + bt^3 + ct^2 + dt + e,\end{equation} where $a > 0$ and $e > 0$.
Then the positivity (non-negativeness) of $P_4(t)$  were proved by Qi-Song-Zhang \cite{QSZ20202}, recently.
\vskip 2mm

(3) $P_4(t) \ge 0$ for all $t\in\mathbb{R}$ if and only if 

\begin{align}& \label{eq:10} \Delta\geq0, \ |b\sqrt{e} - d\sqrt{a} | \le 4\sqrt{ace+2ae\sqrt{ae}} \mbox{ and }\\
	& \label{eq:11}	-2\sqrt{ae} \le c \le 6\sqrt{ae}; \\
	& \label{eq:12} c > 6\sqrt{ae}\mbox{ and  }|b\sqrt{e} + d\sqrt{a}| \le 4\sqrt{ace-2ae\sqrt{ae}},
\end{align}
where 	\begin{equation}\label{eq:13}	 \Delta=4(12ae-3bd+c^2)^3-(72ace+9bcd-2c^3-27ad^2-27b^2e)^2.\end{equation}

(4) $P_4(t) > 0$ for all $t\in\mathbb{R}$ if and only if 	\begin{align}& \label{eq:14}\Delta = 0, b\sqrt{e} = d\sqrt{a}, b^2 +8a\sqrt{ae} = 4ac < 24a\sqrt{ae};\\
	& \label{eq:15}\Delta > 0, |b\sqrt{e} - d\sqrt{a} | \le 4\sqrt{ace+2ae\sqrt{ae}}\mbox{ and }\\
	& \label{eq:16} -2\sqrt{ae} \le c \le 6\sqrt{ae}, \\
	& \label{eq:17}  c > 6\sqrt{ae}\mbox{ and  }|b\sqrt{e} + d\sqrt{a}| \le 4\sqrt{ace-2ae\sqrt{ae}}.
\end{align}
\vskip 4mm

\section{\bf Boundedness from below of scalar potential of two real scalar  and a Higgs boson}
\vskip 4mm
The most general scalar potential of two real scalar fields $\phi_1$ and $\phi_2$ and the Higgs doublet $\mathbf{H}$ (Kannike \cite{K2016,K2018})  is
\begin{equation}\label{eq:18}
	\begin{aligned}
		V(\phi_1,\phi_2,|H|)=&\lambda_{40}\phi_1^4+\lambda_{31}\phi_1^3\phi_2+\lambda_{22}\phi_1^2\phi_2^2+\lambda_{13}\phi_1\phi_2^3+\lambda_{04}\phi_2^4\\
		&+\lambda_{H}|H|^4+\lambda_{H20}|H|^2\phi_1^2+\lambda_{H11}|H|^2\phi_1\phi_2+\lambda_{H02}|H|^2\phi_2^2.
	\end{aligned}
\end{equation}
This scalar potential  defines a 4th order 3 dimensional symmetric tensor $\mathcal{V} = (v_{ijkl})$ with its entries, \begin{equation}\label{eq:19}
	\begin{aligned}v_{1111}=&\lambda_{40},\  v_{2222}=\lambda_{04},\  v_{3333}=\lambda_{H},\ v_{1112}=\frac14\lambda_{31},\ v_{1222}=\frac14\lambda_{13},\\
		v_{1133}=&\frac16\lambda_{H20},\ v_{1122}= \frac16\lambda_{22},\ v_{2233}=\frac16\lambda_{H02},\\
		v_{1233}=&\frac1{12}\lambda_{H11},\ \  v_{ijkl}=0\mbox{ for the others}.
\end{aligned}\end{equation}

In this section, we mainly discuss analytical expressions of positive definiteness of 4th order tensor $\mathcal{V} = (v_{ijkl})$ given by \eqref{eq:19}. Furthermore, we present a necessary and sufficient condition of the boundedness from below of scalar potential of two real scalar  fields $\phi_1$ and $\phi_2$ and the Higgs doublet $\mathbf{H}$.
\vskip 2mm
\subsection{Positive definiteness of 4th order $2$ dimensional symmetric tensors}	
\vskip 2mm

Let $\mathcal{V}=(v_{ijkl})$ be a 4th order $2$ dimensional symmetric tensor with $v_{1111}>0$ and  $v_{2222}>0$. For a vector $x=(x_1,x_2)^\top$ such that
$$\|x\|=\sqrt{x_1^2+x_2^2}=1,$$ we may assume $x_2\not=0$ without loss of generality. We have
\begin{align}\mathcal{V}x^4=&\sum_{i,j,k,l=1}^2v_{ijkl}x_ix_jx_kx_l\nonumber\\
	=&v_{1111}x_1^4+4v_{1112}x_1^3x_2+6v_{1122}x_1^2x_2^2+4v_{1222}x_1x_2^3+v_{2222}x_2^4\nonumber,
\end{align}
and hence,  $$\frac{\mathcal{V}x^4}{x_2^4}
=v_{1111}\left(\frac{x_1}{x_2}\right)^4+4v_{1112}\left(\frac{x_1}{x_2}\right)^3+6v_{1122}\left(\frac{x_1}{x_2}\right)^2+4v_{1222}\left(\frac{x_1}{x_2}\right)+v_{2222}.$$
Clearly, $\mathcal{V}x^4>0$ if and only if $$P(t)=at^4+bt^3+ct^2+dt+e>0, \mbox{ for all } t\in\mathbb{R} $$ with
$$a=v_{1111},\ b=4v_{1112},\ c=6v_{1122},\ d=4v_{1222},\ e=v_{2222}.$$
Then $$\begin{aligned}
	\Delta=&4(12ae-3bd+c^2)^3-(72ace+9bcd-2c^3-27ad^2-27b^2e)^2\\
	=&4(12v_{1111}v_{2222}-48v_{1112}v_{1222}+36v_{1122}^2)^3-(72\times 6v_{1111}v_{1122}v_{2222}\\
	&+72\times 12v_{1112}v_{1122}v_{1222}-72\times 6v_{1122}^3-72\times 6v_{1111}v_{1222}^2\\
	&-72\times 6v_{1112}^2v_{2222})^2\\
	=& 4\times 12^3(I^3-27J^2),
\end{aligned}$$
where $$\begin{aligned}I=&v_{1111}v_{2222}-4v_{1112}v_{1222}+3v_{1221}^2,\\
	J=&v_{1111}v_{1122}v_{2222}+2v_{1112}v_{1122}v_{1222}-v_{1122}^3-v_{1111}v_{1222}^2-v_{1112}^2v_{2222}.
\end{aligned}$$
and hence, the sign of $\Delta$ is the same as one of  $(I^3-27J^2)$.
So, it follows from Eqs. \eqref{eq:14} - \eqref{eq:17} that by simply calculating, 
$\mathcal{V}$ is positive definite, i.e., $\mathcal{V}x^4>0$ for all $x\in\mathbb{R}^2$ if and only if  
$$\begin{cases}
	I^3-27J^2=0,\ \ v_{1112}\sqrt{v_{2222}}=v_{1222}\sqrt{v_{1111}},\\
	v_{1112}^2+2v_{1111}\sqrt{v_{1111}v_{2222}}=6v_{1111}v_{1122}<6v_{1111}\sqrt{v_{1111}v_{2222}};\\
	I^3-27J^2>0,\\
	|v_{1112}\sqrt{v_{2222}}-v_{1222}\sqrt{v_{1111}}|\leq \sqrt{6v_{1111}v_{1221}v_{2222}+2\sqrt{(v_{1111}v_{2222})^3}},\\
	(i) \ -\sqrt{v_{1111}v_{2222}}\leq 3v_{1221}\leq 3\sqrt{v_{1111}v_{2222}};\\
	(ii)\  v_{1221} >\sqrt{v_{1111}v_{2222}}\ \mbox{ and } \\ \ \ |v_{1112}\sqrt{v_{2222}}+v_{1222}\sqrt{v_{1111}}|\leq \sqrt{6v_{1111}v_{1221}v_{2222}-2\sqrt{(v_{1111}v_{2222})^3}}.
\end{cases}\leqno{(\textbf{I})}$$

Similarly, it follows from Eqs. \eqref{eq:10} - \eqref{eq:12} that
$\mathcal{V}=(v_{ijkl})$ is positive semi-definite, i.e.,  $\mathcal{V}x^4\geq0$ for all $x\in\mathbb{R}^2$ if and only if
$$\begin{cases}
	I^3-27J^2\ge0,\\
	|v_{1112}\sqrt{v_{2222}}-v_{1222}\sqrt{v_{1111}}|\leq \sqrt{6v_{1111}v_{1221}v_{2222}+2\sqrt{(v_{1111}v_{2222})^3}},\\
	(i) \ -\sqrt{v_{1111}v_{2222}}\leq 3v_{1221}\leq 3\sqrt{v_{1111}v_{2222}};\\
	(ii) 	\ v_{1221} >\sqrt{v_{1111}v_{2222}}\mbox{ and }\\
	|v_{1112}\sqrt{v_{2222}}+v_{1222}\sqrt{v_{1111}}|\leq \sqrt{6v_{1111}v_{1221}v_{2222}-2\sqrt{(v_{1111}v_{2222})^3}}.
\end{cases}\leqno{(\textbf{II})}$$

Next we give an analytically necessary and sufficient condition of  the boundedness from below of scalar potential of two real scalar  fields $\phi_1$ and $\phi_2$.
The most general scalar potential of two real scalar fields $\phi_1$ and $\phi_2$ may be written as (Kannike \cite{K2016,K2018,K2012}) 
\begin{equation}\label{eq:20} \bar{V}(\phi_1,\phi_2)=\lambda_{40}\phi_1^4+\lambda_{31}\phi_1^3\phi_1+\lambda_{22}\phi_1^2\phi_2^2+\lambda_{13}\phi_1\phi_2^3+\lambda_{04}\phi_2^4.
\end{equation}
Let $\mathcal{V}=(v_{ijkl})$ is the coupling tensor with its entries \begin{equation}\label{eq:21}v_{1111}=\lambda_{40},\  v_{2222}=\lambda_{04},\ v_{1112}=\frac14\lambda_{31},\  v_{1122}=\frac16\lambda_{22},\
	v_{1222}=\frac14\lambda_{13}.\end{equation} In fact, the boundedness from below of two real scalar fields $\phi_1$ and $\phi_2$ is equivalent to  the positive definitnness of the coupling tensor $\mathcal{V}=(v_{ijkl})$. 
Then we have \begin{equation}\label{eq:22}
	\begin{aligned}
		\Delta'=&4(12\lambda_{40}\lambda_{04}-3\lambda_{31}\lambda_{13}+\lambda_{22}^2)^3\\&-(72\lambda_{40}\lambda_{22}\lambda_{04}+9\lambda_{31}\lambda_{22}\lambda_{31} -2\lambda_{22}^3-27\lambda_{40}\lambda_{13}^2-27\lambda_{31}^2\lambda_{04})^2.
	\end{aligned}
\end{equation}
Then from Conditions (\textbf{I}) and (\textbf{II}), the following results are easy to obtain.

Let $\lambda_{40}>0,\ \lambda_{04}>0 $. Then $\bar{V}(\phi_1,\phi_2)>0$  if and only if
$$	\begin{cases}
	\Delta'=0,\ \
	\lambda_{31}\sqrt{\lambda_{04}}=\lambda_{13}\sqrt{\lambda_{40}},\\ \lambda_{31}^2+8\lambda_{40}\sqrt{\lambda_{40}\lambda_{04}}=4\lambda_{40}\lambda_{22}<24\lambda_{40}\sqrt{\lambda_{40}\lambda_{04}};\\
	\Delta'>0,\\
	|\lambda_{31}\sqrt{\lambda_{04}}-\lambda_{13}\sqrt{\lambda_{40}}|\leq 4\sqrt{\lambda_{40}\lambda_{22}\lambda_{04}+2\lambda_{40}\lambda_{04}\sqrt{\lambda_{40}\lambda_{04}}},\\
	(i) \ -2\sqrt{\lambda_{40}\lambda_{04}}\leq \lambda_{22}\leq 6\sqrt{\lambda_{40}\lambda_{04}};\\
	(ii) \ \lambda_{22}> 6\sqrt{\lambda_{40}\lambda_{04}} \mbox{ and }\\ |\lambda_{31}\sqrt{\lambda_{04}}+\lambda_{13}\sqrt{\lambda_{40}}|\leq 4\sqrt{\lambda_{40}\lambda_{22}\lambda_{04}-2\lambda_{40}\lambda_{04}\sqrt{\lambda_{40}\lambda_{04}}}.
\end{cases}\leqno{(\textbf{III})}$$

$\bar{V}(\phi_1,\phi_2)\geq0$  if and only if
$$	\begin{cases}
	\Delta'\geq0,\\
	|\lambda_{31}\sqrt{\lambda_{04}}-\lambda_{13}\sqrt{\lambda_{40}}|\leq 4\sqrt{\lambda_{40}\lambda_{22}\lambda_{04}+2\lambda_{40}\lambda_{04}\sqrt{\lambda_{40}\lambda_{04}}},\\
	(i) \ -2\sqrt{\lambda_{40}\lambda_{04}}\leq \lambda_{22}\leq 6\sqrt{\lambda_{40}\lambda_{04}};\\
	(ii) \ \lambda_{22}> 6\sqrt{\lambda_{40}\lambda_{04}} \mbox{ and }\\ |\lambda_{31}\sqrt{\lambda_{04}}+\lambda_{13}\sqrt{\lambda_{40}}|\leq 4\sqrt{\lambda_{40}\lambda_{22}\lambda_{04}-2\lambda_{40}\lambda_{04}\sqrt{\lambda_{40}\lambda_{04}}}.
\end{cases}\leqno{(\textbf{IV})}$$

In fact, the analytical condition (\textbf{III}) are the boundedness from below in the stronger sense  for the scalar potential \eqref{eq:20} of two real scalar fields $\phi_1$ and $\phi_1$. The analytical condition  (\textbf{IV}) are the analytical boundedness from below condition.

\vskip 2mm
\subsection{Boundedness from below of two real scalar  and a Higgs boson}
\vskip 2mm
The most general scalar potential of two real scalar fields $\phi_1$
and $\phi_2$ and the Higgs doublet $\mathbf{H}$ (Kannike \cite{K2016,K2018})  is
\begin{align}
	V(\phi_1,\phi_2,|H|)=& \lambda_{H}|H|^4+\lambda_{H20}|H|^2\phi_1^2+\lambda_{H11}|H|^2\phi_1\phi_2+\lambda_{H02}|H|^2\phi_2^2\nonumber\\
	&\ +\lambda_{40}\phi_1^4+\lambda_{31}\phi_1^3\phi_2+\lambda_{22}\phi_1^2\phi_2^2+\lambda_{13}\phi_1\phi_2^3+\lambda_{04}\phi_2^4,\label{eq:23}\\
	=& \lambda_{H}|H|^4+M(\phi_1,\phi_2)|H|^2+\bar{V}(\phi_1,\phi_2),  \label{eq:24}
\end{align}
where \begin{equation}\label{eq:25}M(\phi_1,\phi_2)=\lambda_{H20}\phi_1^2+\lambda_{H11}\phi_1\phi_2+\lambda_{H02}\phi_2^2\end{equation} and \begin{equation}\label{eq:26}\bar{V}(\phi_1,\phi_2)=V(\phi_1,\phi_2,0)=\lambda_{40}\phi_1^4+\lambda_{31}\phi_1^3\phi_2+\lambda_{22}\phi_1^2\phi_2^2+\lambda_{13}\phi_1\phi_2^3+\lambda_{04}\phi_2^4.\end{equation}
Recently, Kannike \cite{K2016,K2018} studied the boundedness from below of $V(\phi_1,\phi_2,|H|)$, and gave a  analytical  condition of $V(\phi_1,\phi_2,|H|)>0$.

In this subsection, we will present the analytic  conditions of positive (semi-)definiteness for this special 4th order 3  dimensional symmetric tensor \eqref{eq:19}, and moreover,  the analytic necessary and sufficient conditions are showed for the boundedness from below of scalar potential of two real scalar  fields $\phi_1$ and $\phi_2$ and the Higgs doublet $\mathbf{H}$.

Let $x=(\phi_1,\phi_2,|H|)^\top$. Then  $V(\phi_1,\phi_2,|H|)=\mathcal{V}x^4$, where $\mathcal{V}=(v_{ijkl})$ is a 4th order 3  dimensional symmetric tensor given by \eqref{eq:19}.
Clearly, the tensor given by $\bar{V}(\phi_1,\phi_2)$ is a 4th order 2 dimensional  principal sub-tensor of $\mathcal{V}$.

Let $\lambda_{H}>0$.  It follows from the equation \eqref{eq:24} that
$$\mathcal{V}x^4=\lambda_{H}|H|^4+M(\phi_1,\phi_2)|H|^2+\bar{V}(\phi_1,\phi_2).$$
Which  may be regarded as a quadratic polynomial with respect to $t=|H|^2$, $$P_2(t)=at^2+bt+c, $$  where $$a=\lambda_{H},\ b= M(\phi_1,\phi_2),\ c=\bar{V}(\phi_1,\phi_2).$$ So from Eqs. \eqref{eq:7} and \eqref{eq:8}, it yields to the following conclusion.

$V(\phi_1,\phi_2,|H|)=\mathcal{V}x^4>0$ for all $\phi_1, \phi_2, \mathbf{H}$ if and only if  for all $\phi_1, \phi_2,$
$$	\begin{cases}
	M(\phi_1,\phi_2)\geq0\mbox{ and  }\bar{V}(\phi_1,\phi_2)>0;\\
	M(\phi_1,\phi_2)<0\mbox{  and }4\lambda_{H}\bar{V}(\phi_1,\phi_2)-(M(\phi_1,\phi_2))^2>0.
\end{cases}\leqno{(\textbf{V})}$$

$V(\phi_1,\phi_2,|H|)=\mathcal{V}x^4\geq0$ for all $\phi_1, \phi_2, \mathbf{H}$ if and only if  for all $\phi_1, \phi_2,$
$$	\begin{cases}
	M(\phi_1,\phi_2)\geq0\mbox{ and  }\bar{V}(\phi_1,\phi_2)\ge0;\\
	M(\phi_1,\phi_2)<0\mbox{  and }4\lambda_{H}\bar{V}(\phi_1,\phi_2)-(M(\phi_1,\phi_2))^2\ge0,
\end{cases}\leqno{(\textbf{VI})}$$

It is obvious that $M(\phi_1,\phi_2)=\lambda_{H20}\phi_1^2+\lambda_{H11}\phi_1\phi_2+\lambda_{H02}\phi_2^2$ is a quadric form with respect to two variables $\phi_1,\phi_2$, and hence, the inequality $M(\phi_1,\phi_2)\geq 0$ is equivalent to positive semi-definiteness of its coefficient matrix, $$\left(\begin{matrix}    \lambda_{H20}&  \dfrac12\lambda_{H11}\\
	\dfrac12\lambda_{H11}&  \lambda_{H02}\end{matrix}\right),$$ which is equivalent to
\begin{equation}\label{eq:27}
	\lambda_{H20}\geq0,\ \lambda_{H02}\geq0,\ \lambda_{H20}\lambda_{H02}-\frac14\lambda_{H11}^2\geq 0.
\end{equation}
Similarly, the inequality $M(\phi_1,\phi_2)< 0$ is equivalent to negative definiteness of its coefficient matrix, i.e., the matrix $$\left(\begin{matrix}   - \lambda_{H20}&  -\dfrac12\lambda_{H11}\\
-	\dfrac12\lambda_{H11}& -\lambda_{H02}\end{matrix}\right)$$ is positive definite if and only if 
\begin{equation}\label{eq:28}\lambda_{H20}<0,\ \lambda_{H02}<0,\ \lambda_{H20}\lambda_{H02}-\frac14\lambda_{H11}^2> 0.\end{equation}
At the same time, the inequality $\bar{V}(\phi_1,\phi_2)>0$ can be obtained by the condition (\textbf{III}), and $\bar{V}(\phi_1,\phi_2)\geq0$ can be obtained by the condition (\textbf{IV}). Next we only need show $$4\lambda_{H}\bar{V}(\phi_1,\phi_2)-(M(\phi_1,\phi_2))^2\ge0\ (>0) \mbox{ for all }\phi_1, \phi_2.$$ In order to proving this inequality holds, we take
\begin{equation}\label{eq:29}
	\begin{cases}\lambda_{40}'=4\lambda_{40}\lambda_H-\lambda_{H20}^2,\  \lambda_{04}'=4\lambda_{04}\lambda_H-\lambda_{H02}^2,\\
		\lambda_{31}'=4\lambda_H\lambda_{31}-2\lambda_{H20}\lambda_{H11},\ \lambda_{13}'=4\lambda_H\lambda_{13}
		-2\lambda_{H02}\lambda_{H11},\\
		\lambda_{22}'=4\lambda_H\lambda_{22}-2\lambda_{H20}\lambda_{H02}-\lambda_{H11}^2,\\
		\Delta''=4(12\lambda_{40}'\lambda_{04}'-3\lambda_{31}'\lambda_{13}'+\lambda_{22}'^2)^3\\ \ \ \ \ \ \ \ \ \ \ -(72\lambda_{40}'\lambda_{22}'\lambda_{04}'+9\lambda_{31}'\lambda_{22}'\lambda_{31}' -2\lambda_{22}'^3-27\lambda_{40}'\lambda_{13}'^2-27\lambda_{31}'^2\lambda_{04}')^2.
\end{cases}\end{equation}
Let $ V'(\phi_1,\phi_2)=4\lambda_{H}\bar{V}(\phi_1,\phi_2)-(M(\phi_1,\phi_2))^2$. We may expand the polynomial $V'(\phi_1,\phi_2)$ as follow,
$$\begin{aligned}
	V'(\phi_1,\phi_2)=&4\lambda_{H}\bar{V}(\phi_1,\phi_2)-(M(\phi_1,\phi_2))^2\\
	=&(4\lambda_{40}\lambda_H-\lambda_{H20}^2)\phi_1^4+(4\lambda_H\lambda_{31}-2\lambda_{H20}\lambda_{H11})\phi_1^3\phi_2\\
	&+(4\lambda_H\lambda_{22}-2\lambda_{H20}\lambda_{H02}-\lambda_{H11}^2)\phi_1^2\phi_2^2\\
	&+(4\lambda_H\lambda_{13}-2\lambda_{H02}\lambda_{H11})\phi_1\phi_2^3+(4\lambda_{04}\lambda_H-\lambda_{H02}^2) \phi_2^4\\
	=&\lambda_{40}'\phi_1^4+\lambda_{31}'\phi_1^3\phi_2+\lambda_{22}'\phi_1^2\phi_2^2+\lambda_{13}'\phi_1\phi_2^3
	+\lambda_{04}'\phi_2^4.
\end{aligned}$$
So this definite a 4th order 2 dimensional symmetric tensor $\mathcal{V}=(v_{ijkl})$ with its entries $$v_{1111}=\lambda_{40}',\  v_{2222}=\lambda_{04}',\ v_{1112}=\frac14\lambda_{31}',\  v_{1122}=\frac16\lambda_{22}',\
v_{1222}=\frac14\lambda_{13}'.$$
Let $\lambda_{40}'>0,\ \lambda_{04}'>0 $.   From the condition (\textbf{I}) or (\textbf{III}), we easily obtain the following conclusions. 
\vskip 2mm
$ V'(\phi_1,\phi_2)=4\lambda_{H}\bar{V}(\phi_1,\phi_2)-(M(\phi_1,\phi_2))^2>0$  if and only if
$$\begin{cases}
	\Delta''=0,
	\lambda_{31}'\sqrt{\lambda_{04}'}=\lambda_{13}'\sqrt{\lambda_{40}'},\\ \lambda_{31}'^2+8\lambda_{40}'\sqrt{\lambda_{40}'\lambda_{04}'}=4\lambda_{40}'\lambda_{22}'<24\lambda_{40}'\sqrt{\lambda_{40}'\lambda_{04}'};\\
	\Delta''>0,\\
	|\lambda_{31}'\sqrt{\lambda_{04}'}-\lambda_{13}'\sqrt{\lambda_{40}'}|\leq 4\sqrt{\lambda_{40}'\lambda_{22}'\lambda_{04}'+2\lambda_{40}'\lambda_{04}'\sqrt{\lambda_{40}'\lambda_{04}'}},\\
	(i) \ -2\sqrt{\lambda_{40}'\lambda_{04}'}\leq \lambda_{22}'\leq 6\sqrt{\lambda_{40}'\lambda_{04}'};\\
	(ii) \ \lambda_{22}'> 6\sqrt{\lambda_{40}'\lambda_{04}'}\mbox{ and }\\
	|\lambda_{31}'\sqrt{\lambda_{04}'}+\lambda_{13}'\sqrt{\lambda_{40}'}|\leq 4\sqrt{\lambda_{40}'\lambda_{22}'\lambda_{04}'-2\lambda_{40}'\lambda_{04}'\sqrt{\lambda_{40}'\lambda_{04}'}}.
\end{cases}\leqno{(\textbf{VII})}$$

$ V'(\phi_1,\phi_2)=4\lambda_{H}\bar{V}(\phi_1,\phi_2)-(M(\phi_1,\phi_2))^2\geq0$  if and only if
$$\begin{cases}
	\Delta''\geq0,\\
	|\lambda_{31}'\sqrt{\lambda_{04}'}-\lambda_{13}'\sqrt{\lambda_{40}'}|\leq 4\sqrt{\lambda_{40}'\lambda_{22}'\lambda_{04}'+2\lambda_{40}'\lambda_{04}'\sqrt{\lambda_{40}'\lambda_{04}'}},\\
	(i) \ -2\sqrt{\lambda_{40}'\lambda_{04}'}\leq \lambda_{22}'\leq 6\sqrt{\lambda_{40}'\lambda_{04}'};\\
	(ii) \ \lambda_{22}'> 6\sqrt{\lambda_{40}'\lambda_{04}'}\mbox{ and }\\
	|\lambda_{31}'\sqrt{\lambda_{04}'}+\lambda_{13}'\sqrt{\lambda_{40}'}|\leq 4\sqrt{\lambda_{40}'\lambda_{22}'\lambda_{04}'-2\lambda_{40}'\lambda_{04}'\sqrt{\lambda_{40}'\lambda_{04}'}}.
\end{cases}\leqno{(\textbf{VIII})}$$

Altogether, combing the conditions (\textbf{III}), (\textbf{V}), (\textbf{VII}) and Eqs. \eqref{eq:27}-\eqref{eq:28},  the analytical  necessary and sufficient condition is established for the boundedness from below in the stronger sense of scalar potential of two real scalar  fields $\phi_1$ and $\phi_2$ and the Higgs doublet $\mathbf{H}$.
Let $\lambda_{H}>0$, $\lambda_{40}>0$ and $\lambda_{04}>0$. Then
$V(\phi_1,\phi_2,|\mathbf{H}|)>0$ for all $\phi_1, \phi_2, \mathbf{H}$ if and only if 
$$\begin{cases}
	(1)\ \lambda_{H20}\geq0,\ \lambda_{H02}\geq0,\ 4\lambda_{H20}\lambda_{H02}-\lambda_{H11}^2\geq 0\\
	\Delta'=0,
	\lambda_{31}\sqrt{\lambda_{04}}=\lambda_{13}\sqrt{\lambda_{40}},\\ \lambda_{31}^2+8\lambda_{40}\sqrt{\lambda_{40}\lambda_{04}}=4\lambda_{40}\lambda_{22}<24\lambda_{40}\sqrt{\lambda_{40}\lambda_{04}};\\
	\Delta'>0,\\
	|\lambda_{31}\sqrt{\lambda_{04}}-\lambda_{13}\sqrt{\lambda_{40}}|\leq 4\sqrt{\lambda_{40}\lambda_{22}\lambda_{04}+2\lambda_{40}\lambda_{04}\sqrt{\lambda_{40}\lambda_{04}}},\\
	\ \ -2\sqrt{\lambda_{40}\lambda_{04}}\leq \lambda_{22}\leq 6\sqrt{\lambda_{40}\lambda_{04}};\\
	\ \ \lambda_{22}> 6\sqrt{\lambda_{40}\lambda_{04}}\mbox{ and }\\  |\lambda_{31}\sqrt{\lambda_{04}}+\lambda_{13}\sqrt{\lambda_{40}}|\leq 4\sqrt{\lambda_{40}\lambda_{22}\lambda_{04}-2\lambda_{40}\lambda_{04}\sqrt{\lambda_{40}\lambda_{04}}}.\\
	(2)\ \lambda_{H20}<0,\ \lambda_{H02}<0,\ 4\lambda_{H20}\lambda_{H02}-\lambda_{H11}^2>0,\\ \lambda_{40}'=4\lambda_{40}\lambda_H-\lambda_{H20}^2>0,\ \lambda_{04}'=4\lambda_{04}\lambda_H-\lambda_{H02}^2>0\mbox{ and}\\
	\Delta''=0,
	\lambda_{31}'\sqrt{\lambda_{04}'}=\lambda_{13}'\sqrt{\lambda_{40}'},\\ \lambda_{31}'^2+8\lambda_{40}'\sqrt{\lambda_{40}'\lambda_{04}'}=4\lambda_{40}'\lambda_{22}'<24\lambda_{40}'\sqrt{\lambda_{40}'\lambda_{04}'};\\
	\Delta''>0, \\
	|\lambda_{31}'\sqrt{\lambda_{04}'}-\lambda_{13}'\sqrt{\lambda_{40}'}|\leq 4\sqrt{\lambda_{40}'\lambda_{22}'\lambda_{04}'+2\lambda_{40}'\lambda_{04}'\sqrt{\lambda_{40}'\lambda_{04}'}},\\
	\ \ -2\sqrt{\lambda_{40}'\lambda_{04}'}\leq \lambda_{22}'\leq 6\sqrt{\lambda_{40}'\lambda_{04}'};\\
	\  \ \lambda_{22}'> 6\sqrt{\lambda_{40}'\lambda_{04}'} \mbox{ and }\\
	|\lambda_{31}'\sqrt{\lambda_{04}'}+\lambda_{13}'\sqrt{\lambda_{40}'}|\leq 4\sqrt{\lambda_{40}'\lambda_{22}'\lambda_{04}'-2\lambda_{40}'\lambda_{04}'\sqrt{\lambda_{40}'\lambda_{04}'}}.
\end{cases}\leqno{(\textbf{IV})}$$

Combing the conditions (\textbf{IV}), (\textbf{VI}), (\textbf{VIII}) and Eqs. \eqref{eq:27}, \eqref{eq:28},  the analytical  necessary and sufficient condition is built for the boundedness from below of such a scalar potential \eqref{eq:18}  also.  Let $\lambda_{H}>0$, $\lambda_{40}>0$ and $\lambda_{04}>0$. Then
$V(\phi_1,\phi_2,|\mathbf{H}|)\geq0$ for all $\phi_1, \phi_2, \mathbf{H}$ if and only if
$$\begin{cases}
	(1)\	\lambda_{H20}\geq0,\ \lambda_{H02}\geq0,\ 4\lambda_{H20}\lambda_{H02}-\lambda_{H11}^2\geq 0,\\
	\Delta'\ge0,\\
	|\lambda_{31}\sqrt{\lambda_{04}}-\lambda_{13}\sqrt{\lambda_{40}}|\leq 4\sqrt{\lambda_{40}\lambda_{22}\lambda_{04}+2\lambda_{40}\lambda_{04}\sqrt{\lambda_{40}\lambda_{04}}},\\
	\ \ -2\sqrt{\lambda_{40}\lambda_{04}}\leq \lambda_{22}\leq 6\sqrt{\lambda_{40}\lambda_{04}};\\
	\ \ \lambda_{22}> 6\sqrt{\lambda_{40}\lambda_{04}}\mbox{ and}\\ |\lambda_{31}\sqrt{\lambda_{04}}+\lambda_{13}\sqrt{\lambda_{40}}|\leq 4\sqrt{\lambda_{40}\lambda_{22}\lambda_{04}-2\lambda_{40}\lambda_{04}\sqrt{\lambda_{40}\lambda_{04}}}.\\		
	(2)\	\lambda_{H20}<0,\ \lambda_{H02}<0,\ 4\lambda_{H20}\lambda_{H02}-\lambda_{H11}^2>0, \lambda_{40}'>0,\ \lambda_{04}'>0,\\
	\Delta''\ge0,\\
	|\lambda_{31}'\sqrt{\lambda_{04}'}-\lambda_{13}'\sqrt{\lambda_{40}'}|\leq 4\sqrt{\lambda_{40}'\lambda_{22}'\lambda_{04}'+2\lambda_{40}'\lambda_{04}'\sqrt{\lambda_{40}'\lambda_{04}'}},\\
	\ \ -2\sqrt{\lambda_{40}'\lambda_{04}'}\leq \lambda_{22}'\leq 6\sqrt{\lambda_{40}'\lambda_{04}'};\\
	\ \ \lambda_{22}'> 6\sqrt{\lambda_{40}'\lambda_{04}'}\mbox{ and  }\\
	|\lambda_{31}'\sqrt{\lambda_{04}'}+\lambda_{13}'\sqrt{\lambda_{40}'}|\leq 4\sqrt{\lambda_{40}'\lambda_{22}'\lambda_{04}'-2\lambda_{40}'\lambda_{04}'\sqrt{\lambda_{40}'\lambda_{04}'}}.
\end{cases}\leqno{(\textbf{V})}$$

\section{Conclusions}

In this paper, for a  scalar potential of two real scalar  fields $\phi_1$ and $\phi_2$ and the Higgs doublet $\mathbf{H}$,  the analytically necessary and sufficient conditions of the boundedness from below  are   achieved  with the help of the analytical expressions of positive definiteness for  4th order 2-dimension symmetric tensors. More precisely,  for a 4th order 2-dimension symmetric tensor, 
\begin{itemize}
	\item The condition (\textbf{I}) is an analytically necessary and sufficient condition of positive  definiteness;
	\item The condition (\textbf{II}) is an analytically necessary and sufficient condition of positive semi-definiteness.
\end{itemize}
For a  scalar potential of two real scalar  fields $\phi_1$ and $\phi_2$,
\begin{itemize}
	\item The condition (\textbf{III}) is an analytically necessary and sufficient condition of the boundedness from below in the stronger sense;
	\item The condition (\textbf{IV}) is an analytically necessary and sufficient condition of the boundedness from below .
\end{itemize}
For a  scalar potential of two real scalar  fields $\phi_1$ and $\phi_2$ and the Higgs doublet $\mathbf{H}$,
\begin{itemize}
	\item The condition (\textbf{V}) is an analytically necessary and sufficient condition of the boundedness from below;
	\item the condition (\textbf{IV}) is an analytically necessary and sufficient condition of the boundedness from below in the stronger sense.
\end{itemize}

\section*{Competing interest}
The authors declared that they have no conflict of interest. 


\end{document}